# Fast Response to Infection Spread and Cyber Attacks on Large-Scale Networks

Sven Leyffer* Ilya Safro†

**Abstract**

We present a strategy for designing fast methods of response to cyber attacks and infection spread on complex weighted networks. In these networks, nodes can be interpreted as primitive elements of the system, and weighted edges reflect the strength of interaction among these elements. The proposed strategy belongs to the family of multiscale methods whose goal is to approximate the system at multiple scales of coarseness and to obtain a solution of microscopic scale by combining the information from coarse scales. In recent years these methods have demonstrated their potential for solving optimization and analysis problems on large-scale networks. We consider an optimization problem that is based on the SIS epidemiological model. The objective is to detect the network nodes that have to be immunized in order to keep a low level of infection in the system.

## 1 Introduction

Networks are a widely used type of abstraction for complex data of scientific interest for which one may want to emphasize the relationships between primitive components of the system (such as nodes and vertices) by connecting them with edges or links. Examples can be found in domains such as astronomy, Internet, food webs, and metabolic networks. Recent growth of large-scale, real-world network data available for scientific analysis has promoted significant theoretical and practical advances in many areas of natural sciences and engineering; for example see [2, 30]. Optimization of different quantitative objectives on networks often plays a crucial role in network science, not only when a practical solution is needed, but also for a general understanding of structural and statistical features of networks.

In recent decades, a significant amount of research in the networks science has been done in analyzing infection spreading. Examples can be found in domains such as epidemiology [16, 21, 28], cyber security [14], and social sciences [2]. Developing strategies of the response policies to the infection propagation is crucial for applications of theoretical analysis in the real-life situations. Usually, such response strategies can be formulated as optimization models that consider applying some operation (e.g., immunization of individuals, rerouting of network links, updating the antivirus software) on network primitives (such as nodes, links, and communities) at different resolutions. Often, such optimization models must consider the number of *limited* resources that prohibit performing *all* possible operations for better response results. For example, when an epidemics occurs it is impossible to immunize everybody immediately. Instead, we have to formulate the policies that can help to achieve a particular goal given partial immunization with limited resources. In this paper we propose a method for designing efficient methods for optimal response problems on large-scale networks. Implementation of the method is available for download at [27].

*Mathematics and Computer Science Division, Argonne National Laboratory, leyffer@mcs.anl.gov
†Mathematics and Computer Science Division, Argonne National Laboratory, safro@mcs.anl.gov

## 2 Optimal Response Model

We consider a traditional infection-spread model in which network nodes can be in one of two possible states, namely, *infected* and *susceptible* (SIS model; see [20]), and each node $i$ is associated with probability of being infected at time $t$, $\phi_{i,t}$. Introduced as a simplification of the SIR (susceptible-infected-recovered) model in [24], the SIS model has been extensively analyzed in epidemiology and adapted in the cyber security area for analysis of computer viruses propagation [23]. In this paper we follow that general model and the probabilistic version of optimal response model (formulated in [17]) that takes into account the status of all individuals in the network.

The SIS model considers the following quantities: $S$ - number of susceptible nodes; $I$ - number of infected nodes; $\beta$ - infection transmission rate; and $\delta$ - rate of recovery from infection. The model describes an evolution of the two classes of population of infected nodes $I$ and susceptible nodes $S$ at time $t$ as

$$\begin{cases} \frac{dI}{dt} = \lambda S - \delta I \\ \frac{dS}{dt} = \delta I - \lambda S, \end{cases} \tag{1}$$

where $\lambda = \beta\langle k\rangle I/(S+I)$ reflects the rate at which susceptible nodes become infected and $\langle k\rangle$ is an average node degree. One of the most important consequences of (1) is the notion of an *epidemic threshold* $\tau$, a measure to predict when the infection outbreak disappears, that is, the value that has to be compared with $\beta/\delta$. Chakrabarti et al. proposed a topology-independent, nonlinear dynamical system model of SIS [9]. Their model,

$$1 - \phi_{i,t} = (1-\phi_{i,t-1})h_{i,t} + \delta\phi_{i,t-1}h_{i,t}, \quad i = 1...n, \tag{2}$$

describes the probability that node $i$ is susceptible, where $h_{i,t}$ is the probability that node $i$ is not infected from its neighbors at time $t$. We assume that the probabilities of nodes being infected in the previous round $t-1$ are independent; see [9] for more details. Often, the infection transfer rate cannot be represented by a single parameter $\beta$; thus, without loss of generality it is replaced by a matrix $P^{n\times n} = \{p_{ij}\}$, where $p_{ij}$ is the probability of node $i$ being infected by node $j$. The probability of an uncompromised node $i$ not being infected at time $t$ is

$$h_{i,t} = \prod_{j\in\Gamma_i} (1 - p_{ij}\phi_{j,t-1}). \tag{3}$$

We now formulate the optimization problem whose goal is to keep the level of infection at each node low while maximizing the (weighted) number of "alive" connections between nodes that will not be considered by the policy as "needing special attention". This is motivated by the infection spread response policies in different domains that are often driven by the number of resources available for the realization. We denote the graph underlying a given network by $G = (V, E, w)$, where $V$ is a set of nodes, $E$ is a set of edges, and $w : V \to \mathbb{R}_{\geq 0}$ is a weighting function on $E$ that represents the strength of connectivity between two nodes (such as the number of shared users between sites in a cyber system). Assuming that the probabilities of infection transition from $\Gamma_i$ (neighbors of $i$) to $i$ are independent, the problem is formulated as

$$\begin{aligned}
\underset{x}{\text{maximize}} \quad & \sum_{ij \in E} w_{ij} x_i x_j \\
\text{subject to} \quad & x_i - \prod_{j \in \Gamma_i} (1 - p_{ij}\phi_{j,t-1} x_j) \leq b_i \quad \forall i \in V, \\
& x_i \in \{0, 1\} \qquad\qquad\qquad\qquad\quad \forall i \in V,
\end{aligned} \tag{4}$$

where $w_{ij}$ is the link weight between nodes $i$ and $j$; $b_i$ is a threshold for bounding the level of allowed probability of infection at node $i$; and $x_i$ are binary variables (1 - if we decide to leave the node $i$ functioning, 0 - closed, requiring special attention). If for some $ij \in E$ one of the nodes $i$ or $j$ is solved to be closed, that link does not contribute its $w_{ij}$ to the objective. In general, (4) is a nonconvex integer nonlinear program and known to be $NP$-complete [17]. There exist a number of deterministic solvers for (4), including BARON [35] based on a branch-and-reduce strategy that employs piecewise linear underestimators of the multilinear functions on 4 to construct a convex relaxation, and then branches either on an integer variable or on a nonconvexity creating a branch-and-bound tree-search. The open-source solver Couenne [3] implements a similar strategy. However, these solvers cannot handle problems with tens of thousands of nodes (typically they work well for problems with a few hundred variables and nonlinear expressions), because the search tree explodes exponentially. Therefore, in this paper, we propose a strategy for designing fast, suboptimal multiscale methods for this class of problems. Such methods are often more useful in practice than are optimal ones that take a long time to converge even for small instances.

## 3 Multiscale Strategy

In many practical situations, it is often noticeable that when elementary parts of a system have a complicated behavior, their ensembles often can be much more structured. The multiscale computational methodology [7, 6] is a systematic approach for achieving efficient calculations of systems containing many degrees of freedom (such as network nodes, image pixels, and particles). From the relationships among the given microscopic parts of the system, the rules for the system at increasingly coarser scales are derived. The idea behind multiscale methods is to collect the relevant information regarding the system at different scales and then to obtain the solution at the microscopic scale by adapting the information inherited from coarse scales. Realizations of the multiscale frameworks are attractive in practice because they can be naturally combined with other computational and analysis techniques, making them suitable for applying on large-scale instances [7]. For many applications with underlying computational structural problems on networks (or graphs), the introduction of multiscale methods has led to breakthroughs in the quality of computational and data analysis results. Examples include structural analysis problems [36, 32], partitioning [19], clustering [13], segmentation [37], VLSI design [29], and linear ordering [34].

Our method is inspired by the *algebraic multigrid* (AMG) [38] applied for solving optimization problems on large-scale graphs [34]. In this framework a hierarchy of decreasing-size network graph Laplacians $\{L_i\}_{i=0}^k$ is created by a process called *coarsening*, starting from the original network $L_0$. When a small-enough (or easy-to-solve) Laplacian $L_k$ is created, the problem is solved exactly for $L_k$; and the solution is projected to the original $L_0$ by interpolating it scale after scale. Each interpolation is then followed by refinement, that is, by local processing that improves the solution (see Figure 1).

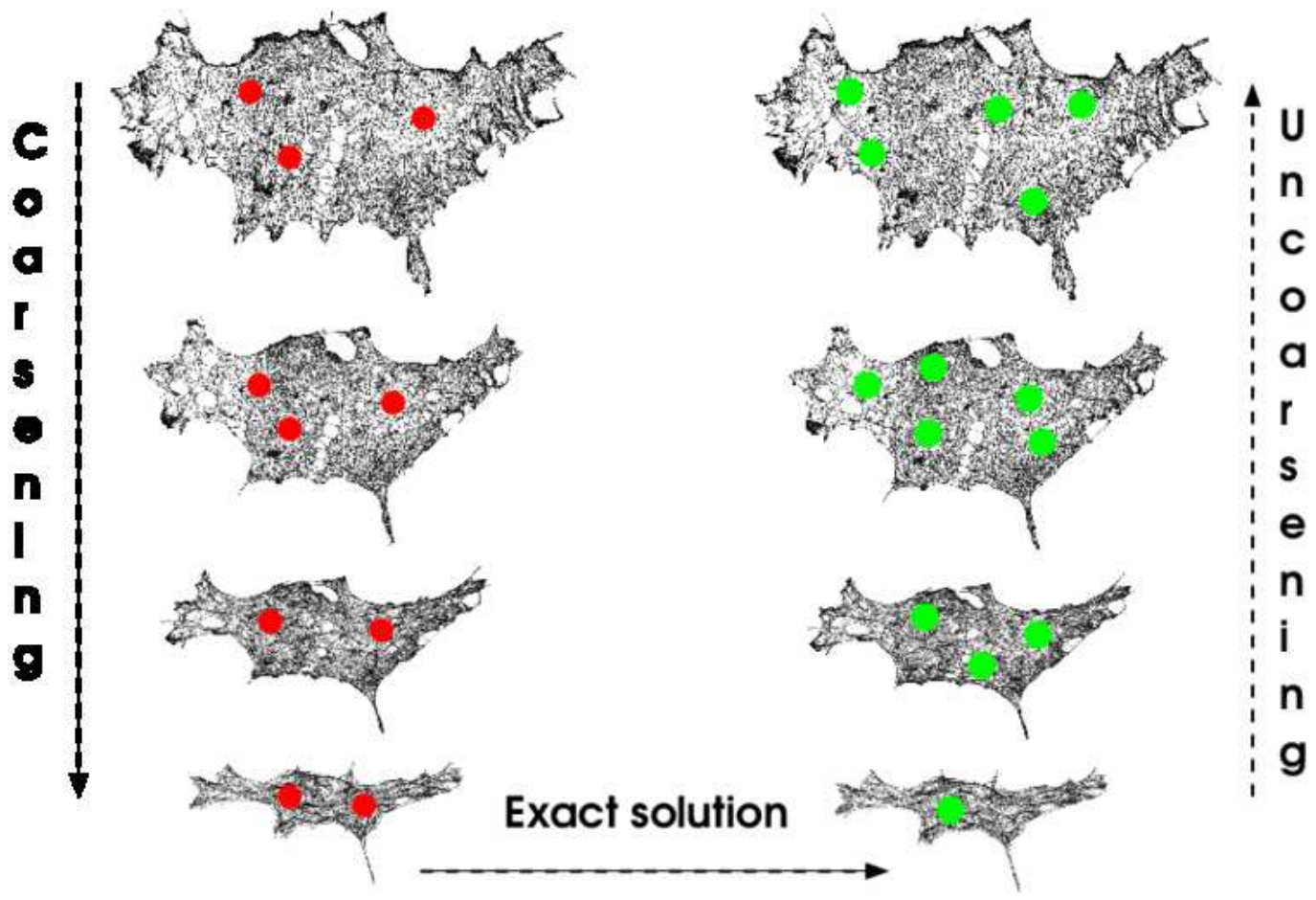

Figure 1: V-cycle scheme for US roads network. Red and green circles illustrate compromised and closed areas, correspondingly.

### 3.1 Coarse problem

The construction of a coarse problem on a network consists of two main phases: defining the sets of coarse nodes, and links. For both phases it is important to be able to describe how "close" two given fine-level nodes are to each other at the stage of switching to the coarse level. In particular, it is crucial in the context of infection spread on real networks when the weights on the links can be noisy and the measure of closeness must take into account the neighborhoods of nodes, instead of looking at one particular direct connection. A recently introduced approach of algebraic distance between nodes [34, 10] has proved itself being successful in AMG-based methods [8]. We define the distance between nodes $i$ and $j$ as

$$\varrho_{ij}^{(k)} := \left( \sum_{r=1}^{R} \left| \chi_i^{(k,r)} - \chi_j^{(k,r)} \right|^p \right)^{1/p} , \tag{5}$$

where the superscript $^{(k,r)}$ refers to the $k$th iteration on the $r$th initial random vector, namely, $\chi^{(k,r)} = H^k \chi^{(0,r)}$, and $H$ is a Jacobi over-relaxation iterator of the Laplacian (see the Appendix). The set of coarse nodes $V_c$ is created by the aggregation of fine nodes $V_f$ into small clusters based on the strength of connectivity estimated by using the algebraic distance. The nodes from $V_f$ are traversed one by one and divided into two sets $C$ and $F$ such that (a) $C \cup F = V_f$; and (b) nodes in $F$ are strongly coupled to $C$ (see the Appendix). Then nodes in $F$ are divided among some of their neighbors in $C$ to form future coarse nodes. The nodes are traversed in the ascending order of the infection's level. By doing so we localize small areas of high connectivity that potentially can propagate the infection rapidly (and, thus, have to be be determined as "closed" in the solution).

The coarse network Laplacian $L_c$ is defined by the restriction operator $L_c = R^T L_f R$, where $R \in \mathbb{R}^{n \times N}$ is a matrix of connections between variables in $F$ and $C$, where $n = |V_f|$ and

$N = |C|$. Finally, the optimization problem is formulated for aggregated (coarse) variables as

$$\begin{aligned} \underset{X}{\text{maximize}} \quad & \sum_{IJ \in E_c} W_{IJ} X_I X_J + \sum_{I \in V_c} A_I X_I \\ \text{subject to} \quad & X_I - \prod_{J \in \Gamma_I} (1 - P_{IJ} \Phi_{J,t-1} X_J) \le B_I \quad \forall I \in V_c, \\ & X_I \in \{0,1\} \quad \forall I \in V_c, \end{aligned} \tag{6}$$

where $X_I$ are binary decision variables that correspond to coarse nodes; $W_{IJ}$ and $P_{IJ}$ are accumulated strengths of connectivity and infection spread probabilities between aggregates $I$ and $J$ in $V_c$, respectively; $\Phi_I$ are infection probabilities for coarse nodes; and $B_I$ are accumulated thresholds for infection level for coarse nodes (see the Appendix). The main difference between the fine and the coarse problems is the new linear term $\sum_{I \in V_c} A_I X_I$. It takes into account the fine-level links between nodes aggregated into the same coarse node.

### 3.2 Uncoarsening

During the coarsening process we recursively form the hierarchy of smaller problems until a small-enough level is reached. The size of this level depends on the external optimization solver one can use (see the Appendix). After the coarsest problem is solved, the solution is gradually projected back to the original scale. It consists of three phases: $C$-nodes interpolation, $F$-nodes interpolation, and refinement. Initially, all seed nodes $i \in C$ are initialized by $x_i = X_I$, where $X_I$ is a corresponding coarse variable seeded by node $i \in C$. Next, all $F$-nodes are interpolated by maximizing their contribution to the current objective. This is equivalent to solving (4) when all $x_j$'s are fixed except the node $i$ that is being currently interpolated. As a result of these two phases we obtain the first feasible solution of the fine-scale problem. This solution is then improved by Gauss-Seidel-like relaxation in which for every node the contribution of the opposite solution to the objective (or/and number of satisfied constraints) is compared with its current contribution.

The refinement phase consists of the collective improvement of the solution for sufficiently small subsets of variables. For this purpose we have formulated a *localized refinement* procedure, which extracts from the entire system small subproblems and solves each separately. We note that this phase can easily be performed in parallel by using for example the red-black order of the refinements [38] (see Figure 2). Single subset refinement solves problem (7) for subset of nodes $S$ by choosing a connected subgraph and fixing the *boundary conditions* for the rest of the nodes. The single localized refinement is formulated as

$$\begin{aligned} \underset{x}{\text{maximize}} \quad & \sum_{i,j \in S} w_{ij} x_i x_j + \sum_{i \in S, j \notin S} w_{ij} x_i \tilde{x}_j + \sum_{i \in S} a_i x_i \\ \text{subject to} \quad & x_i - k_i \prod_{\substack{j \in \Gamma_i \\ j \in S}} (1 - p_{ij} \phi_{j,t-1} x_j) \le b_i \ \forall i \in V, \\ & x_i \in \{0,1\} \ \forall i \in V, \end{aligned} \tag{7}$$

where $\tilde{x}_j$ is a fixed solution for node $j \notin S$, and

$$k_i = \prod_{j \in \Gamma_i, j \notin S} (1 - p_{ij} \phi_{j,t-1} \tilde{x}_j) \, .$$

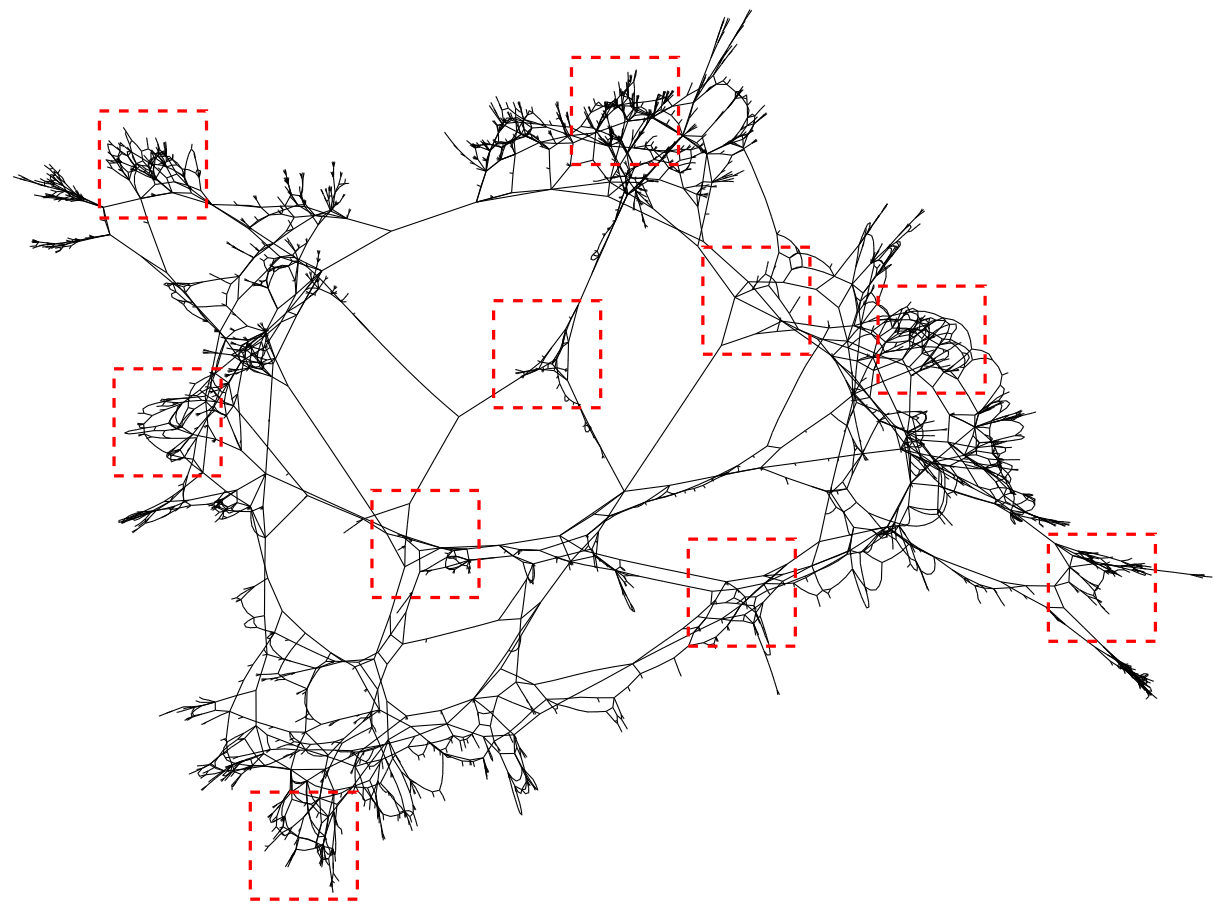

Figure 2: Localized refinement. Red dashed squares correspond to subgraphs induced by small subsets of nodes for localized refinements

# 4 Results

We evaluate our method on a set of small networks with known optimal solutions, two case studies (HIV spread and cyber infrastructure networks), and one massive data set. The two case study networks are typical complex network instances on which solving this particular optimization is of great practical importance. The connection between epidemiological models and analysis of cyber attacks has been extensively investigated during the past two decades [22, 23, 9]. The massive dataset evaluation contains networks of different structures and sources, including some that arise in applications that are not related immediately to the response problem but can potentially represent hard structures for the method.

## 4.1 Networks with Known Optimal Solutions

Before analyzing the proposed method on large-scale instances that cannot be solved to proven optimality (even using commercial solvers) reasonably fast, we evaluated how good the results of the multiscale method are on small networks that can be solved exactly. For this purpose we generated 200 Erdös-Rényi networks [15] with randomly initiated $w_{ij}$ and $\phi_{i,t-1}$ (see (4)). The results are demonstrated in Figure 3. Typically in such settings almost half of the instances are optimally solved while others are close enough to the optimum (between 90% and 100%).

## 4.2 HIV Spread Network

First we demonstrated our algorithm on a network created from data that was collected in by Potterat et al. [31] related to the HIV spread over individuals who were in contact through sex or injection drug use. The original data contains a network with 250 nodes, where each node corresponds to an individual. We generated 100 similar networks by using a multiscale network generator [18] and connected them by several random edges in order to create one big network (see Figure 4). We simulated an immediate outbreak of the infection in which initially 5% of nodes were associated with high level of infection ($\phi_i \in [0.8, 1]$) and each edge had the same rate of infection transmission. Then five iterations of the infection spread were performed; at each iteration, all nodes released their infection to the neighbors, and the updated $\phi$ was

$$\forall i \in V \quad \phi_i^{\text{new}} = \min\left(1, \phi_i^{\text{old}} + \sum_{j\in\Gamma_i} \frac{p_{ij}}{\sum_{k\in\Gamma_i} p_{ik}} \phi_j^{\text{old}}\right).$$

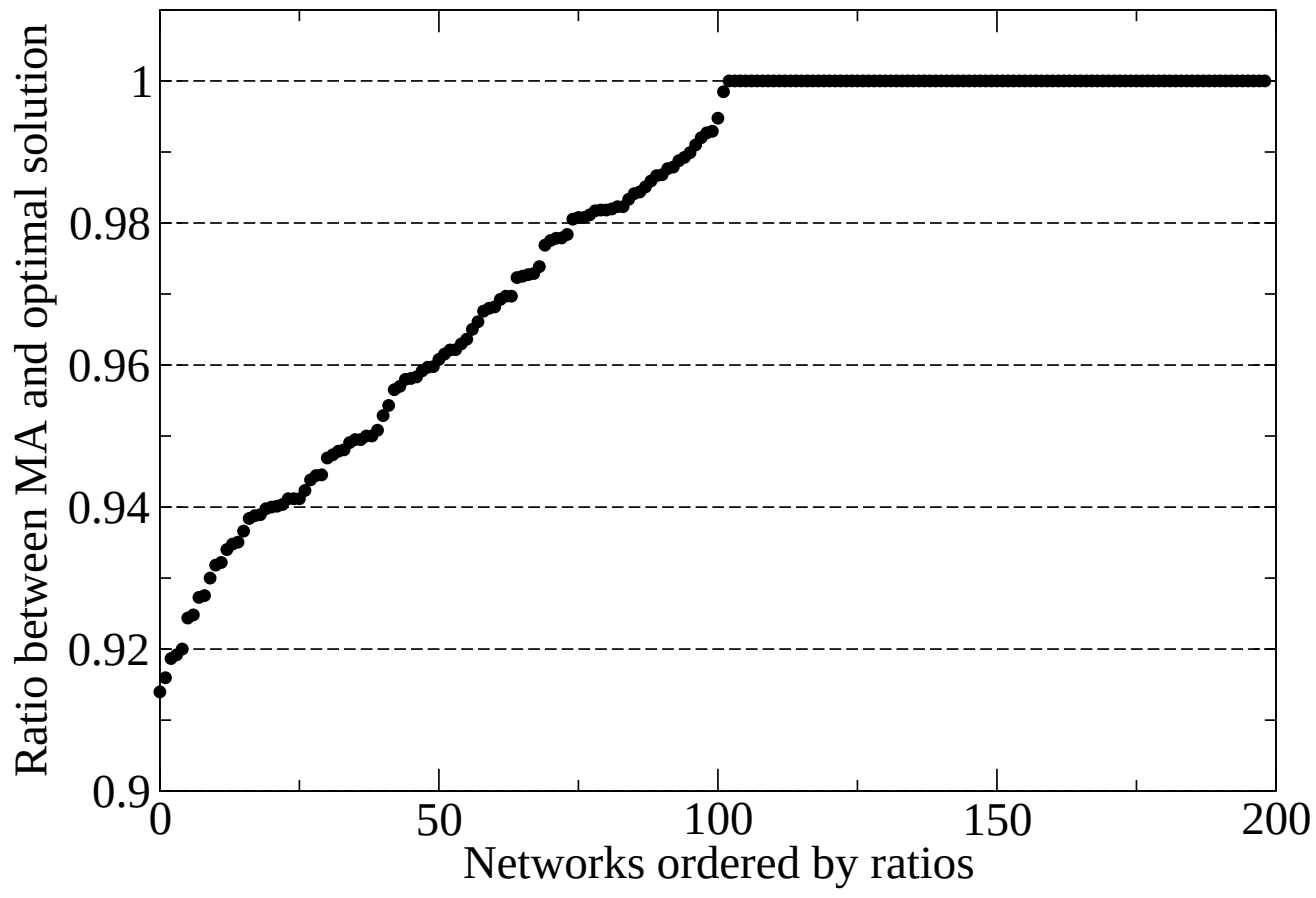

Figure 3: Comparison with optimal solutions for 200 small networks. Each point represents a ratio between the objectives of MA and optimal solutions, respectively, for one network.

Typical computational results comparing the multiscale algorithm (MA) and a combination of different types of iterated local search (ILS) are presented in Figure 5. The multiscale algorithm reached the objective 13404 in just five iterations of the refinement, while ILS was able to achieve the objective 12870 being more than 100 times slower than MA. We note that the content of the two solutions was different. The number of nodes suggested to consider as "closed" by MA (8159) was bigger than those chosen by ILS (7864). In contrast to ILS, MA left "open" more high-degree nodes. We observe that introducing the linear penalty term $-\sum_{i\in V}\sum_{j\in\Gamma_i} w_{ij}x_i$ to the objective of (4) may reverse this situation in favor of closing more high-degree nodes. Such a term can be coarsened similarly to the aggregated edge coefficients $A_i$ in (6).

### 4.3 Peer-to-Peer Network

Peer-to-peer systems (P2P) are a type of technology for collaborative environments in which each participating computer can play roles of both server and client. At the core of such systems lies an infrastructure for sharing computational resources such as storage space and CPU time. Data streams in such networks are often associated with mutually anonymous (for users) source and target nodes which brings the realization of a strong cyber security system to one of the system's central issues. Examples of P2P systems include Napster, Gnutella, and SETI@Home. Altunay et al. analyzed one such system [1], namely, the Open Science Grid, and proposed optimization model for manipulating collaboration policies to prevent the fast spread of cyber attacks. Unfortunately, methods proposed in [1] are too slow for large-scale networks.

We evaluated our method on the biggest connected component of the Gnutella P2P network [25, 33]. As in the previous case we compared our results with those of ILS. We observed that ILS rapidly reaches slow improvement zones; however, in contrast to the previous case there was a significant gap in the objective between MA and ILS on this type of network, and thus the evaluation consists of 30 trials with different random initial seeds. The results are shown in Figure 6. Each bar corresponds to the ratio between MA and ILS objectives for one initial random seed. The difference in running time is similar to the previous case (between 100 and 200 times) in favor of MA. In addition, MA typically finds a better solution, as shown in Figure 6. The difference in solution quality can be as much as 20%.

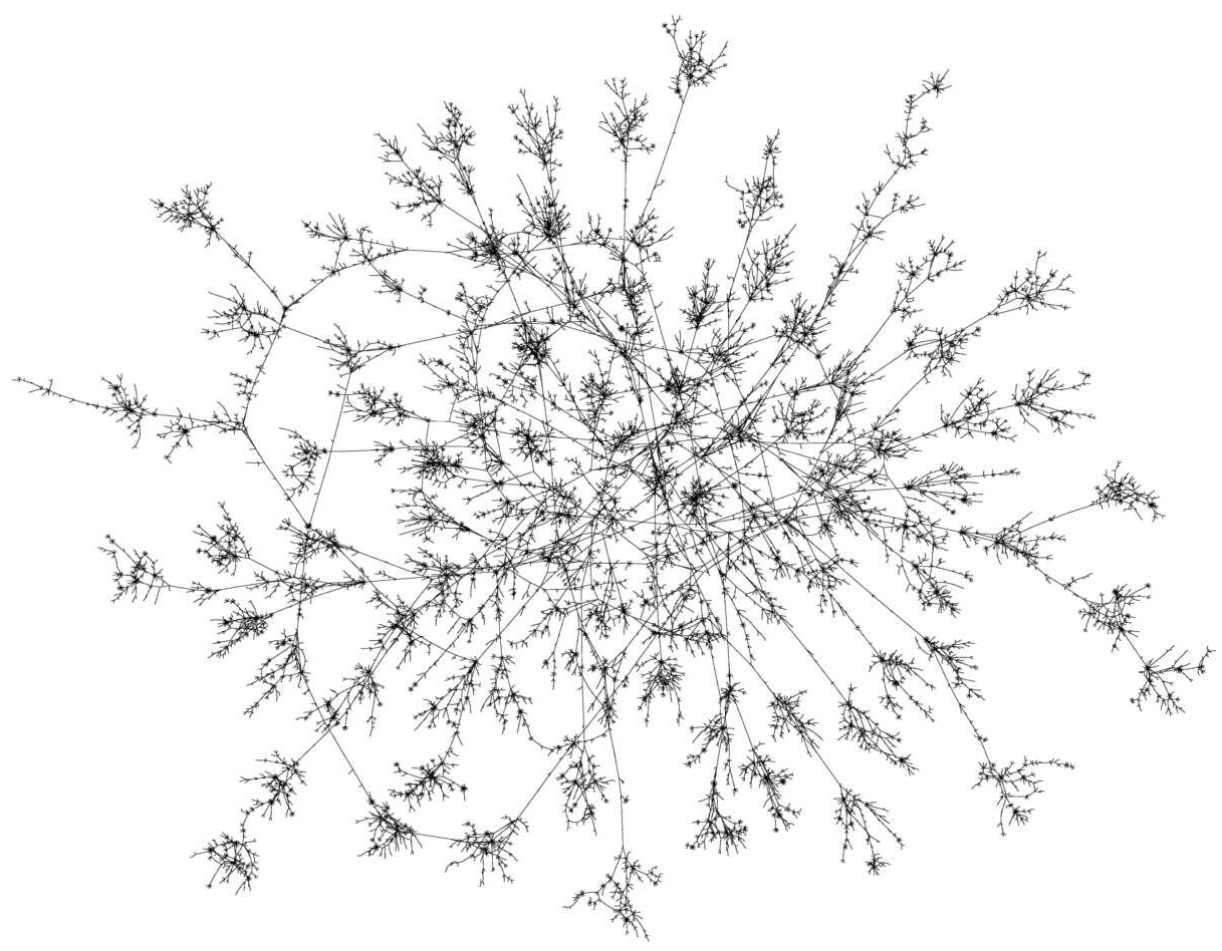

Figure 4: Infection spread network ($|V| = 25090, |E| = 28284$) constructed by sparse random connections among 100 generated networks that are similar to real HIV spread data.

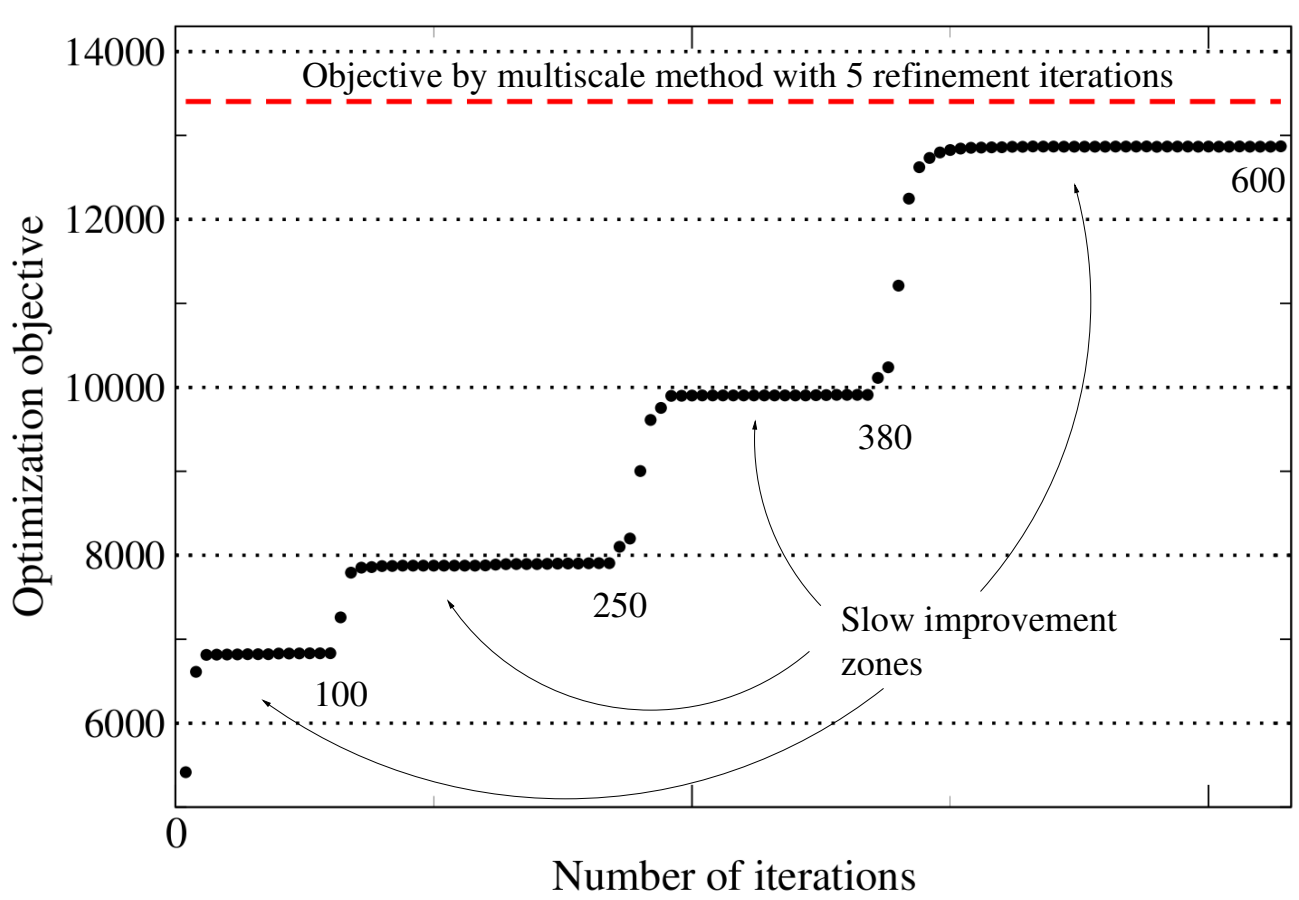

Figure 5: Computational results on the infection spread network.

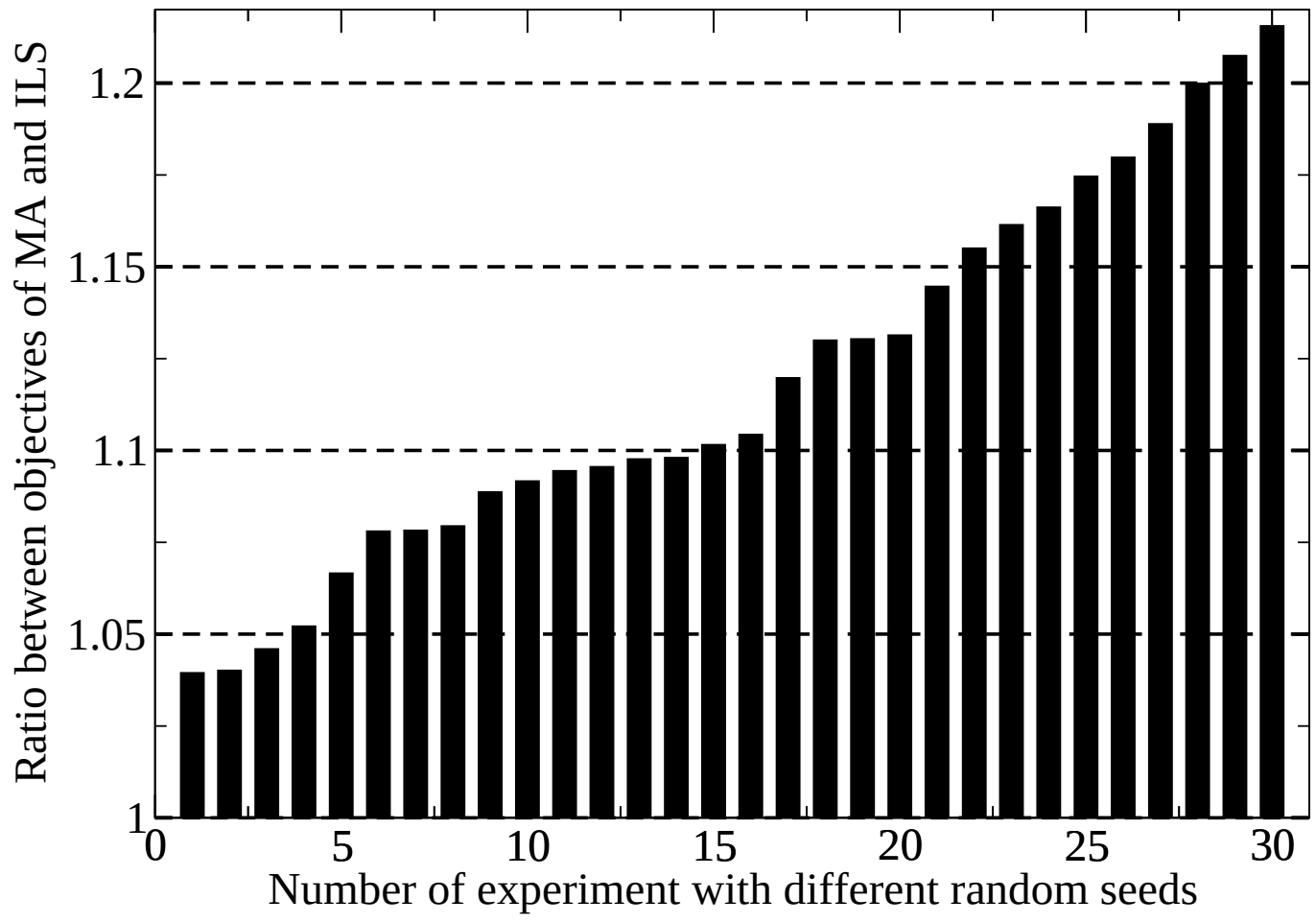

Figure 6: Computational results on the Gnutella P2P network.

### 4.4 Massive Simulations

We also demonstrated the robustness of the proposed method on a test set of 100 large-scale graphs taken from different sources such as [25, 12, 11]. In contrast to HIV and P2P networks, the connection of many of these graphs to the infection spread response problem is not straightforward (if at all), but their structural complexity presents a particular difficulty for optimization methods. The results of comparison with ILS are presented in Figure 7, where each point corresponds to the ratio between objectives of MA and ILS. For approximately one-third of the test set, the difference in the objective is significant while the running time of MA is between 50 to 200 times faster. The difference in running time depends mostly on the size of subproblems (7) because in many cases the external solvers such as [26] that ensure upper bounds are not of linear complexity. We note that according to the results, the most difficult instances are networks with high average degree. The biggest difference detected between MA and ILS ratios was 132 for a graph with average degree 240. We generated 100 graphs by high-entropic multiscale editing [18] and observed that MA still improves the objective over ILS with a factor between 70 and 150.

## 5 Conclusions

We propose a fast multiscale method for optimizing the response policies to infection spread in large-scale, complex, weighted networks. The method is flexible and can be easily adapted for different changes in the model formulation such as changing the model to *link-based* immunization [4, 1] and adding penalty function to the objective and new constraints. Similar to many methods in the large family of multiscale algorithms, our approach is scalable and suitable for parallelization on HPC systems. As key future work directions we identify two branches: theoretical and applied. Theoretical work involves rigorous analysis in order to identify upper and lower bounds. In the applied branch we suggest to introduce similar optimization problems for the SIR model, where the "recovered" states of nodes will be introduced and PDE-based constraints will describe time series of the data.

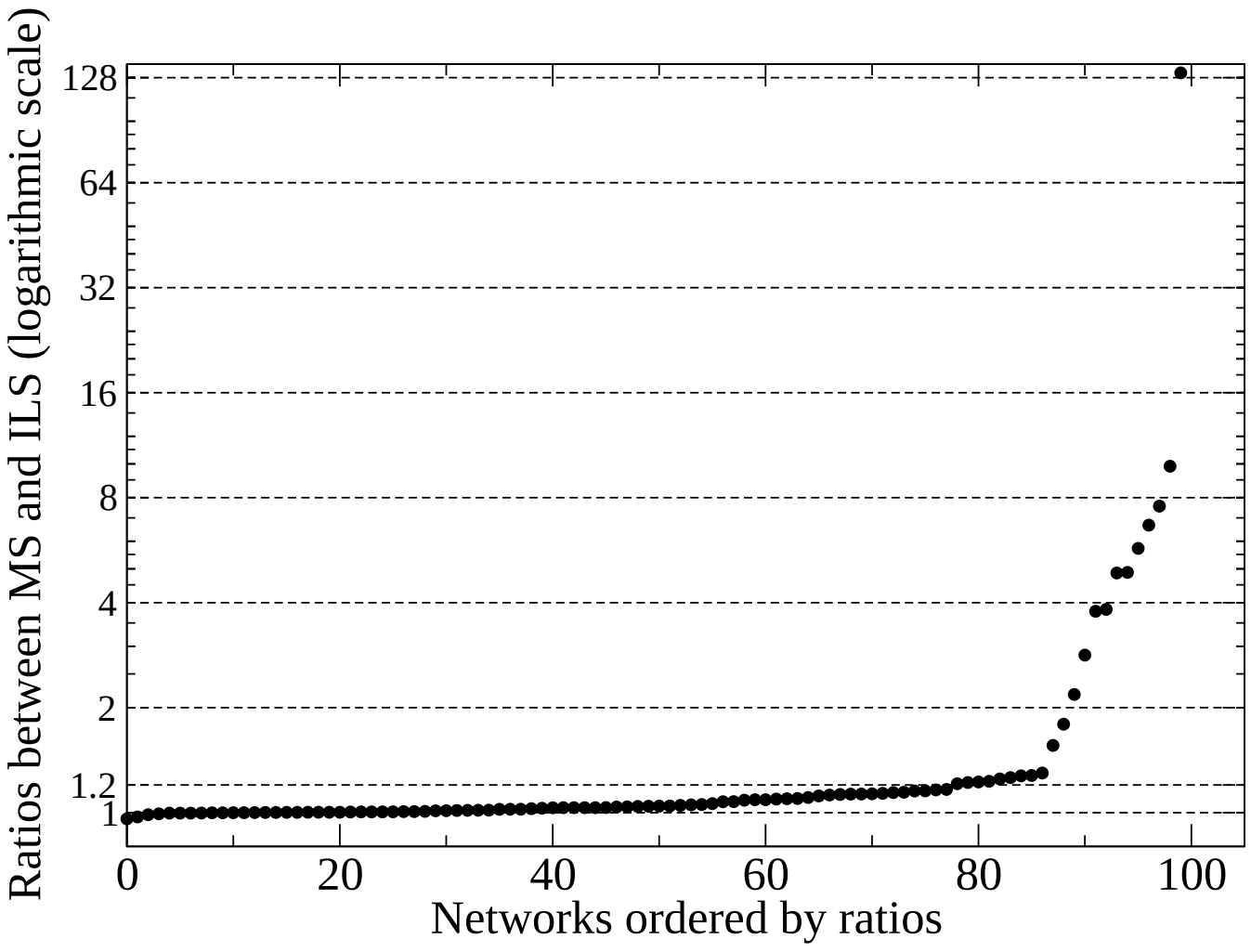

Figure 7: Massive data experiments on various graph structures.

# 6 Appendix: Technical Details

## 6.1 Algebraic Distance

The algebraic distance can be based on different types of stationary iterative relaxations such as Gauss-Seidel and Jacobi [10]. We experimented with the Jacobi overrelaxation iterator $H = (1-\omega)I + \omega D^{-1}W$ in order to validate the ability to make the implementation fully parallel (in contrast to Gauss-Seidel which is difficult to parallelize). Here $I$ is a diagonal matrix with ones on the diagonal, $W$ is a weighted adjacency matrix with entries $w_{ij}$, and $D$ is a diagonal matrix with entries $d_{ii} = \sum_{j \in V} w_{ij}$.

## 6.2 $F - C$ Coupling and Restriction Operator

The selection of nodes to $C$ is done by traversing all nodes starting with those that are highly infected. Initially we set $F = V_f$ and $C = \emptyset$. Then the node is added to $C$ if $\sum_{j \in C} \rho_{ij}^{-1} / \sum_{j \in V_f} \rho_{ij}^{-1} \geq \Theta$. The default value for parameter $\Theta$ is 0.5. Increasing it will usually lead to slower coarsening and potentially better results, as more scales are created during the coarsening and more refinement is done. Decreasing $\Theta$ will lead to the opposite effects.

## 6.3 Aggregated Variables and Constants

Coefficients $A_I$ in (6) accumulate weights of edges $w_{ij}$ whose endpoints are aggregated into $I$, namely, $A_I = \sum_{i,j \in I} w_{ij}$. Coarse nodes represent small clusters of fine nodes and, thus, edge weights between coarse nodes are $W_{IJ} = \sum_{k \neq l} R_{kI} w_{kl} R_{lJ}$. In order to reduce the complexity of the coarse problems the number of non-zeros in rows of $R$ has to be bounded by a sufficiently small number. In the experiments we determined that one nonzero entry with the strongest algebraic distance coupling [34] is enough for practical purposes. Coefficients $P_{IJ}$ can be derived similarly if $p_{ij}$ are not derived from $w_{ij}$. In our experimental settings they are normalized inverses of the given connection strengths. Aggregation of scalars $\phi_{j,t-1}$ and $b_I$ into $\Phi_{J,t-1}$ and $B_I$, respectively, has to be done according to the application, because in some situations the probability of infection in coarse node $J$ may not depend linearly on those in fine-level node. In our simplified models they are accumulated from the corresponding fine level nodes as normalized sums.

### 6.4 External Optimization Solver

The recently developed mixed-integer nonlinear optimization toolkit MINOTAUR [26] has proved itself as particularly suitable for such problems. MINOTAUR compares favorably with other state-of-the-art MINLP solvers such as BONMIN [5]. We used MINOTAUR as both a coarsest level and local-processing solver for the problems in (7). The sizes of the coarsest level and local processing problems were 40 and 15 variables, respectively. During our experiments all small problems of the type (7) were solved exactly.